\documentclass{PoS}
\usepackage[latin1]{inputenc}
\usepackage{amsmath}
\usepackage{booktabs}

\newcommand{\be}{\begin{equation}}
\newcommand{\ee}{\end{equation}}

\title{The kaon mass in $\mathbf{N_f=2+1+1}$ flavor twisted mass Wilson ChPT}

\ShortTitle{The kaon mass in 2+1+1 flavor twisted mass Wilson ChPT}

\author{\speaker{Oliver B\"ar}\\
       Institute of Physics\\
        Humboldt University Berlin\\
	12489 Berlin\\
	Germany \\
       E-mail: \email{obaer@physik.hu-berlin.de}}

\author{Ben H\"orz\\
        Institute of Physics\\
        Humboldt University Berlin\\
	12489 Berlin\\
	Germany\\
       E-mail: \email{hoerz@physik.hu-berlin.de}}

\abstract{We construct the chiral low-energy effective theory for $N_f =2+1+1$ flavor lattice QCD with twisted mass Wilson fermions. In contrast to existing results we assume a heavy charm quark mass such that the D mesons are too heavy to appear as degrees of freedom in the effective theory. As an application we compute the kaon mass to 1-loop order in the LCE regime. The result contains a chiral logarithm involving the neutral pion mass which has no analogue in continuum ChPT.}

\FullConference{31st International Symposium on Lattice Field Theory LATTICE 2013\\
		 July 29 - August 3, 2013\\
		 Mainz, Germany}

\begin{document}

\section{Introduction}
Lattice simulations with twisted mass Wilson fermions offer various advantages, most prominently O($a$) improvement at maximal twist \cite{Frezzotti:2003ni}. One of the disadvantages is the explicit breaking of parity and flavor symmetry, which leads to a mass splitting between the charged and neutral pion. This splitting vanishes in the continuum limit  and is not a fundamental concern. However, the splitting is rather large at the nonzero lattice spacings in the simulations. Table \ref{table1} summarizes data published in \cite{Herdoiza:2013sla} for ensembles at three different lattice spacings, ranging from approximately 0.09fm (A lattices) to 0.06fm (D lattices). The neutral pion is significantly lighter than the charged one, with a mass that can be half the mass of the charged one for the coarsest lattices. For the finest lattice spacing the splitting is much less severe. Still, the neutral pion mass is roughly 20 percent lighter than the charged one.

 Some possible consequences of such pion mass splitting have been discussed in Ref.\ \cite{Bar:2010jk}. Within twisted mass Wilson chiral perturbation theory (tm WChPT)  \cite{Sharpe:1998xm,Sharpe:2004ny} the pion masses and the pion decay constant were computed to 1-loop order, taking the large pion mass splitting at leading order (LO) in the chiral expansion. The results differ from the corresponding ones in continuum ChPT. Chiral logarithms with the neutral pion mass appear, which can impact the chiral extrapolation to the physical point. Moreover, one expects finite volume (FV) corrections which are significantly enhanced for small neutral pion masses \cite{Colangelo:2010cu}. The last two columns of table \ref{table1} display the combination $M_{\pi}L$, both for the charged and the neutral pion masses. A common rule of thumb says that the FV corrections are small for $M_{\pi}L\sim 4$ or larger. Most of the ensembles in table \ref{table1} satisfy this criterion with the charged pion mass. However, the situation looks much less comfortable with the neutral pion mass. In this case twelve of the seventeen ensembles have values $M_{\pi^0}L$ less than 4, some of them even less than 2.
Facing these numbers one can expect significant FV corrections which are completely underestimated with the continuum ChPT results.

With the findings of \cite{Bar:2010jk} in mind one may ask about the impact of the pion mass splitting on other observables, and here we report our result for the kaon mass.

\begin{table}[tbh]
\begin{center}
		\begin{tabular}{c c c c c c}
			\toprule
			Ensemble & $a M_{\pi^\pm}$ & $a M_{\pi^0}$ & $M_{\pi^0} / M_{\pi^\pm}$ & $M_{\pi^\pm} L$ & $M_{\pi^0} L$ \\
			\midrule \midrule
			A30.32&0.1234&0.0611&0.50&3.9&2.0 \\
			A40.32&0.1415&0.0811&0.57&4.5&2.6 \\
			A40.24&0.1445&0.0694&0.48&3.5&1.7 \\
			A60.24&0.1727&0.1009&0.58&4.1&2.4 \\
			A80.24&0.1987&0.1222&0.61&4.8&2.9 \\
			A100.24&0.2215&0.1570&0.71&5.3&3.8 \\
			A80.24s&0.1982&0.1512&0.76&4.8&3.6 \\
			A100.24s&0.2215&0.1863&0.84&5.3&4.5 \\\hline
			B25.32&0.1064&0.0605&0.57&3.4&1.9 \\
			B35.32&0.1249&0.071&0.57&4.0&2.3 \\
			B55.32&0.154&0.1323&0.86&4.9&4.2 \\
			B75.32&0.1808&0.1557&0.86&5.8&5.0 \\
			B85.24&0.1931&0.1879&0.97&4.6&4.5 \\\hline
			D15.48&0.0695&0.0561&0.81&3.3&2.7 \\
			D20.48&0.0797&0.0651&0.82&3.8&3.1 \\
			D30.48&0.0978&0.086&0.88&4.7&4.1 \\
			D45.32sc&0.1198&0.0886&0.74&3.8&2.8 \\
			\bottomrule
		\end{tabular}
		\end{center}
		\caption{Charged and neutral pion masses for various ensembles generated by the ETM collaboration. Data from Ref.\ \cite{Herdoiza:2013sla}. Central values only are given, for errors see \cite{Herdoiza:2013sla}.}
		\label{table1}
	\end{table}

\section{Charmless ChPT for $\mathbf{N_f =2+1+1}$ flavor QCD}
$N_{f}=2+1+1$ tm lattice QCD as used by the ETM collaboration involves four Wilson quark flavors with a mass term of the form \cite{Frezzotti:2003xj}
\be\label{FourFlavorMassterm}
M = 
\left(
\begin{array}{cc}
M_l & 0 \\
0 & M_h
\end{array}
\right)\,,
\ee
where the 2-by-2 submatrices $M_{l,h}$ are the mass matrices in the light and heavy sector,
\be
M_l  =m + i \sigma_3\gamma_5 \mu_l\,,\qquad M_h  =  m + i \sigma_1\gamma_5 \mu_h + \delta \sigma_3\,.
\ee
The untwisted mass parameter $m$ is tuned to achieve maximal twist. The charged pion mass is essentially determined by the light twisted mass $\mu_l$, while the masses of the heavier kaons and $D$ mesons are related to $\mu_h$ and $\delta$. 

The main point here is the trivial fact that twisted mass fermions necessarily come in pairs. In particular, a strange quark is always accompanied by  its ``twist partner'', the charm quark. This may not be regarded a disadvantage. Present day lattice simulations have reached a precision where one may be sensitive to dynamical charm quark effects, at least in some observables. However, the construction of chiral perturbation theory for quarks with the mass term in (\ref{FourFlavorMassterm}) is not straightforward.  The familiar procedure to set up the low-energy effective theory is to promote the mass term (and any other symmetry breaking terms) to a spurion field with definite transformation properties under various symmetry transformations. In particular, the transformations involving flavor symmetry involve the full heavy flavor doublet, and not only the strange quark. So the standard procedure treats the $D$ and $D_s$ mesons as (pseudo) Goldstone bosons just as the pions and kaons. Of course, this is justified only in the case of (unphysical) light charm quark masses. 

The 4-flavor ChPT with mass degenerate kaons and $D$ mesons ($\delta=0$ in $M_h$) has been constructed long ago in \cite{AbdelRehim:2006ve}. The extension to $\delta \neq0$ has been studied more recently in \cite{Munster:2011gh}. No matter how useful these papers have been (for example to study automatic O($a$) improvement at maximal twist), the results, in particular those for the kaon mass, are not applicable to the phenomenologically relevant case with $M_D\gg M_K$. 

The question is how to construct ChPT without treating the $D$ mesons as pseudo Goldstone bosons (``charmless ChPT'', for short)?
One way to proceed is illustrated by the following well known example. Consider standard continuum SU(3) ChPT \cite{Gasser:1984gg}
and suppose we are interested in energies and momenta much below the kaon mass. In this case the kaon and eta meson degrees of freedom decouple and one must reproduce the results of SU(2) ChPT. Explicitly one proceeds as follows: In the standard SU(3) ChPT field we drop the kaon and $\eta$ fields,
\be\label{SU3Fields}
\Sigma_{(3)} = \exp\left(\frac{2i}{f_{(3)}} \pi\right)\,, \qquad\pi = \frac{1}{\sqrt{2}} \left(
\begin{array}{ccc}
\frac{\pi^0}{\sqrt{2}} & \pi^+ &0\\
\pi^- & -\frac{\pi^0}{\sqrt{2}} & 0\\
0 & 0& 0
\end{array} \right)\,.
\ee
Here we have written the field $\Sigma$ and the leading order low-energy constant (LEC) $f$ with a subscript ``$(3)$'', making explicit the number of flavors. Plugging the reduced field (\ref{SU3Fields}) into the SU(3) chiral lagrangian we do not quite reproduce the SU(2) chiral lagrangian, since the strange quark mass still appears explicitly. However, after absorbing this mass dependence in various LECs we obtain the familiar SU(2) chiral lagrangian, For example, the $L_4$ term in \cite{Gasser:1984gg} leads to the relation
\be
f_{(2)} = f_{(3)} \left(1+8\frac{B_{(3)} m_s}{f^2_{(3)}}L_4\right)
\ee
between the LECs in the 2-flavor and 3-flavor theories, making the leading analytic $m_s$ dependence of $f_{(2)}$ explicit. After this absorption we have completely undone the expansion in the strange quark mass and end up with ChPT for the two light quark flavors only.

We can follow the same strategy to construct charmless twisted mass Wilson ChPT (tm WChPT). Our starting point is the chiral lagrangian for $N_f=2+1+1$ tm WChPT given in \cite{Munster:2011gh}. We drop the heavy $D$ and $D_s$ meson fields, insert the reduced chiral field into the chiral lagrangian, and absorb the remaining explicit $m_c$ dependence in the various LECs. In complete analogy we find 
\be
f_{(3)} = f_{(4)} \left(1+8\frac{B_{(4)} m_c}{f^2_{(4)}}L_4\right)\,.
\ee
In addition, some O($am_c$) terms are absorbed in the LECs associated with the nonzero lattice spacing effects of Refs.\ \cite{Rupak:2002sm,Bar:2003mh}, e.g.
\be
W_{0,(3)} = W_{0,(4)} \left(1+8\frac{B_{(4)} m_c}{f^2_{(4)}}W_6\right)
\ee
The reduction to the 3-flavor chiral lagrangian is not as simple as sketched here, because the identification of the heavy {\em physical} fields is not trivial with twisted mass terms. It requires finding the nontrivial vacuum state by solving a gap equation which itself depends on the charm quark mass. Nevertheless, in the end one obtains a reduced 3-flavor chiral lagrangian without any explicit reference to $m_c$. The expansion in the charm quark mass, which would be very poor for physical charm quark masses, is completely undone and for practical calculations one can simply forget  the 4-flavor theory as a starting point. Still, the construction via the 4-flavor theory guarantees that we captured correctly all possible terms in the reduced 3-flavor theory,  since the 4-flavor chiral lagrangian contained all terms allowed by the symmetries of $N_f=2+1+1$ tm lattice QCD.     

\section{Application: The kaon mass in the LCE regime}
The reduced 3-flavor chiral lagrangian and details about its construction will be presented elsewhere \cite{BarHoerz}.
In the following we will show a straightforward application, namely the calculation of the kaon mass in the LCE regime.\footnote{This regime is sometimes called Aoki regime. For a description of the order counting see Ref.\cite{Aoki:2008gy}, for example.}  This regime assumes a sufficiently small light quark mass $\mu_{l} \sim a^2$. In this case the O($a^2$) terms in the chiral lagrangian are taken at LO, just as the kinetic and the mass terms, ${\cal L}_{\rm LO} = {\cal L}_2 + {\cal L}_{a^2}$. This regime is the appropriate one if the pion mass splitting is of the order of the pion masses itself. We only consider the case of maximal twist in both the light and the heavy sector. This is the relevant case in practice and the calculation is significantly simpler than for non-maximal twist angles, since all terms proportional to the cosines of the twist angles vanish.

The tree level masses of the pion and kaon fields are trivially obtained by expanding the reduced lagrangian to quadratic order:
\begin{eqnarray}
&&M_{\pm}^2 =2B\mu_{l}\,,\nonumber\\
&&M_{0}^2 =2B\mu_{l} +2c_2a^2\,,\label{LOMasses}\\
&& M_{K}^2 =B(\mu_{l} +\mu_h + \delta)\,.\nonumber
\end{eqnarray}
$M_{\pm}$ and $M_{0}$ denotes the charged and neutral pion mass, respectively.
$M_K$ is the LO mass for all four kaons, which are degenerate for the twisted mass term considered here \cite{Chiarappa:2006ae}. The pion mass splitting $\Delta M^2_{\pi,{\rm LO}}=M_0^2 - M_{\pm}^2$ shows up at LO by construction. We have written the splitting in terms of $c_2$, which is proportional to the combination $2W_6'+W_8'$ of LECs in ${\cal L}_{a^2}$.

The NLO computation of $M_{\rm K}$ follows the one for the pion mass in Ref.\ \cite{Bar:2010jk}. Key features of this calculation are: (i) One has to keep track of the different pion masses in the pion propagators, and (ii) additional vertices of O($a^2$) need to be taken into account that stem from ${\cal L}_{a^2}$. Our final result for the kaon mass reads
\begin{eqnarray}
M^2_{K,{\rm NLO}} &=& M^2_{K} 
\left[1+\frac{1}{48\pi^2}\frac{M^2_{\eta}}{f^2}\ln\frac{M_{\eta}^2}{\mu^2}
+8L_{46}\frac{M^2_{\pm}}{f^2}
+8(2L_{46}+L_{58})\frac{M^2_{K}}{f^2}
\right]\nonumber\\
&&+\,\,\hat{a}^2\, W^{\prime}_{78}
\frac{M^2_K}{4\pi^2f^4}\ln \frac{M^2_K}{\mu^2}-\,\,\Delta M^2_{\pi}  
\frac{M^2_0}{64\pi^2f^2}\ln \frac{M^2_0}{\mu^2} + {\rm analytic}\,.\label{MKNLO}
\end{eqnarray}
The result is expressed in terms of the LO masses given in eq.\ (\ref{LOMasses}). The first line on the right hand side of (\ref{MKNLO})
 is the familiar result of Gasser and Leutwyler \cite{Gasser:1984gg}, expressed in terms of LEC combinations $L_{ij}=2L_j-L_i$. Note, however, that it is the charged pion mass that appears in the contribution proportional to $L_{46}$. The second line contains the additional contributions proportional to the lattice spacing. All these terms vanish for $a\rightarrow 0$, hence the Gasser-Leutwyler result is recovered in the continuum limit. For nonzero lattice spacings, on the other hand, two additional chiral logs contribute, and neither of them involves $M_{\eta}$ as the continuum contribution. The chiral log with the kaon mass is proportional to the LEC combination $W^{\prime}_{78}=2W_7'+W_8'$. Estimates for $W_8'$ exist \cite{Herdoiza:2013sla} but to date nothing is known about the size of $W_7'$. In contrast, the prefactor of the chiral log involving the neutral pion mass is proportional to $2W_6'+W_8'$, which is nothing but the pion mass splitting. So this non-analytic contribution is essentially determined by the pion masses. 
This is particularly beneficial since the pionic chiral log is prone to produce sizable FV corrections. 

How serious the impact of the O($a^2$) contributions to the kaon mass is needs to be checked by analyzing actual lattice data.
However, a rough estimate for the pion log contribution can be given. We define the relative shift of the kaon mass by the modulus of
\be\label{DeltaMK}
\frac{\Delta M_K}{M_K} = \frac{\Delta M^2_{\pi}}{2M^2_K}  
\frac{M^2_{\pi^0}}{64\pi^2f_{\pi}^2}\left[\ln \frac{M^2_{\pi^0}}{\mu^2}+\tilde{g}_1(M_{\pi^0}L)\right]\,,
\ee
where $\tilde{g}_1$ denotes the usual FV correction stemming from a finite spatial volume with periodic boundary conditions (see \cite{Colangelo:2003hf}, for example). On the right hand side we may use the measured values in \cite{Herdoiza:2013sla,Baron:2010bv} for the various quantities involved. The results $\Delta M_K/M_K$ (for $\mu=1$GeV) are listed in table 2. We also listed the statistical error for the kaon mass, which is at the sub percent level. The upshot of this naive comparison is that both uncertainties are comparable, and the pionic chiral log correction should not be discarded straight away. 
\begin{table}[t]
\begin{center}
		\begin{tabular}{c c c c c c c c}
			\toprule
			Ensemble & $a M_{\pi^\pm}$ & $a M_{\pi^0}$ & $aM_K$ & $a f_{\pi}$ & $L / a$ & $\frac{\Delta M_K}{M_K} \left [ \% \right ]$ & $\frac{ \Delta M_{K,\text{stat}}}{M_K}  \left [ \% \right ]$ \\
			\midrule \midrule
			A30.32&0.1234&0.0611&0.25150(29)&0.046&32&0.02&0.12 \\
			A40.32&0.1415&0.0811&0.25666(23)&0.048&32&0.12&0.09 \\
			A40.24&0.1445&0.0694&0.25884(43)&0.046&24&0.10&0.16 \\
			A60.24&0.1727&0.1009&0.26695(52)&0.051&24&0.14&0.19 \\
			A80.24&0.1987&0.1222&0.27706(61)&0.054&24&0.27&0.22 \\
			A100.24&0.2215&0.1570&0.28807(34)&0.056&24&0.37&0.12 \\
			A80.24s&0.1982&0.1512&&&24&& \\
			A100.24s&0.2215&0.1863&0.26502(90)&0.055&24&0.32&0.34 \\\hline
			B25.32&0.1064&0.0605&0.21240(50)&0.040&32&0.003&0.24 \\
			B35.32&0.1249&0.0710&0.21840(28)&0.043&32&0.09&0.13 \\
			B55.32&0.1540&0.1323&0.22799(34)&0.046&32&0.17&0.15 \\
			B75.32&0.1808&0.1557&0.23753(32)&0.049&32&0.23&0.13 \\
			B85.24&0.1931&0.1879&0.24476(44)&0.049&24&0.06&0.18 \\
			\bottomrule
		\end{tabular}
		\end{center}
		\caption{Data for pseudo scalar masses and the pion decay constant used to estimate $\Delta M_{K}/M_K$ (see text). 
		Data from Refs.\ \cite{Herdoiza:2013sla} and \cite{Baron:2010bv}. 
		Central values only except for the kaon mass, which includes the statistical error. }
		\label{fig2}
	\end{table}

\section{Summary and outlook}
Current twisted mass lattice QCD simulations show a sizable pion mass splitting due to explicit flavor symmetry breaking. Twisted mass Wilson ChPT provides (modified) ChPT formulae which can be used to assess the impact of large pion mass splitting on the chiral extrapolation and on FV corrections caused by small neutral pion masses. Two main results have been reported here. Firstly, a charmless chiral lagrangian for $N_f=2+1+1$ tm lattice QCD has been constructed, which removes in a consistent manner the heavy degrees of freedom associated with a heavy charm quark.  Secondly, based on this lagrangian the kaon mass has been computed to 1-loop order in the LCE regime. As anticipated, additional chiral logarithms proportional to $a^2$ show up at this order. In particular, a chiral log involving the neutral pion mass is present in the result, which is expected to be the dominant source for FV corrections.

A natural next step is the computation of the kaon decay constant. It requires the charmless  expression for the physical axial vector current, which can be derived following the steps for the construction of the charmless effective lagrangian.

\section*{Acknowledgements}
OB is supported in part by the Deutsche Forschungsgemeinschaft (SFB/TR 09).


\begin{thebibliography}{99}

\bibitem{Frezzotti:2003ni}
  R.~Frezzotti and G.~C.~Rossi,
  JHEP {\bf 0408} (2004) 007
  [hep-lat/0306014].
  
\bibitem{Herdoiza:2013sla}
  G.~Herdoiza, K.~Jansen, C.~Michael, K.~Ottnad and C.~Urbach,
  JHEP {\bf 1305} (2013) 038
  [arXiv:1303.3516 [hep-lat]].
  
\bibitem{Bar:2010jk}
  O.~B\"ar,
  Phys.\ Rev.\ D {\bf 82} (2010) 094505
  [arXiv:1008.0784 [hep-lat]].
  
\bibitem{Sharpe:1998xm}
  S.~R.~Sharpe and R.~L.~Singleton, Jr,
  Phys.\ Rev.\ D {\bf 58} (1998) 074501
  [hep-lat/9804028].
  
\bibitem{Sharpe:2004ny}
  S.~R.~Sharpe and J.~M.~S.~Wu,
  Phys.\ Rev.\ D {\bf 71} (2005) 074501
  [hep-lat/0411021].
  
\bibitem{Colangelo:2010cu}
  G.~Colangelo, U.~Wenger and J.~M.~S.~Wu,
  Phys.\ Rev.\ D {\bf 82} (2010) 034502
  [arXiv:1003.0847 [hep-lat]].

\bibitem{Frezzotti:2003xj}
  R.~Frezzotti and G.~C.~Rossi,
  Nucl.\ Phys.\ Proc.\ Suppl.\  {\bf 128} (2004) 193
  [hep-lat/0311008].
  
\bibitem{AbdelRehim:2006ve}
  A.~M.~Abdel-Rehim, R.~Lewis, R.~M.~Woloshyn and J.~M.~S.~Wu,
  Phys.\ Rev.\ D {\bf 74} (2006) 014507
  [hep-lat/0601036].
  
\bibitem{Munster:2011gh}
  G.~M\"unster and T.~Sudmann,
  JHEP {\bf 1104} (2011) 116
  [arXiv:1103.1494 [hep-lat]].

\bibitem{Gasser:1984gg}
  J.~Gasser and H.~Leutwyler,
  Nucl.\ Phys.\ B {\bf 250} (1985) 465.

\bibitem{Rupak:2002sm}
  G.~Rupak and N.~Shoresh,
  Phys.\ Rev.\ D {\bf 66} (2002) 054503
  [hep-lat/0201019].
  
\bibitem{Bar:2003mh}
  O.~B\"ar, G.~Rupak and N.~Shoresh,
  Phys.\ Rev.\ D {\bf 70} (2004) 034508
  [hep-lat/0306021].
  
\bibitem{BarHoerz}
O.~B\"ar and B.~H\"orz, in preparation.

\bibitem{Aoki:2008gy}
  S.~Aoki, O.~B\"ar and B.~Biedermann,
  Phys.\ Rev.\ D {\bf 78} (2008) 114501
  [arXiv:0806.4863 [hep-lat]].

\bibitem{Chiarappa:2006ae}
  T.~Chiarappa, F.~Farchioni, K.~Jansen, I.~Montvay, E.~E.~Scholz, L.~Scorzato, T.~Sudmann and C.~Urbach,
  Eur.\ Phys.\ J.\ C {\bf 50} (2007) 373
  [hep-lat/0606011].

  
\bibitem{Colangelo:2003hf}
  G.~Colangelo and S.~D\"urr,
  Eur.\ Phys.\ J.\ C {\bf 33} (2004) 543
  [hep-lat/0311023].
 
  
\bibitem{Baron:2010bv}
  R.~Baron, P.~.Boucaud, J.~Carbonell, A.~Deuzeman, V.~Drach, F.~Farchioni, V.~Gimenez and G.~Herdoiza {\it et al.},
  JHEP {\bf 1006} (2010) 111
  [arXiv:1004.5284 [hep-lat]].
  


\end{thebibliography}
\end{document}